\documentclass[showpacs]{revtex4}
\usepackage{epsfig}
\def\beq{\begin{equation}}
\def\enq{\end{equation}}
\def\beqa{\begin{eqnarray}}
\def\enqa{\end{eqnarray}}

\def\GeV{\nobreak\,\mbox{GeV}}

\def\mixs{\lag\bar{s}g\si.Gs\rag}

\def\G3{\lag g^3G^3\rag}

\def\si{\sigma}

\def\al{\alpha}
\def\be{\beta}
\def\alma{\alpha_{max}}
\def\almi{\alpha_{min}}
\def\bemi{\beta_{min}}
\def\lb{\label}

\newcommand{\rag}{\rangle}
\newcommand{\lag}{\langle}

\begin{document}

\title{\sc
Erratum: QCD sum rules study of the $J^{PC}=1^{--}$ charmonium $Y$ 
mesons}
\author{R.M. Albuquerque}
\email{rma@if.usp.br}
\affiliation{Instituto de F\'{\i}sica, Universidade de S\~{a}o Paulo,
C.P. 66318, 05389-970 S\~{a}o Paulo, SP, Brazil}
\author{M. Nielsen}
\email{mnielsen@if.usp.br}
\affiliation{Instituto de F\'{\i}sica, Universidade de S\~{a}o Paulo,
C.P. 66318, 05389-970 S\~{a}o Paulo, SP, Brazil}

\begin{abstract}
We correct a mistake in the analytical expression given in
 Nucl. Phys. {\bf A} 815, 53 (2009) [arXiv:0804.4817]
for the $D_{s0}\bar{D}_s^*$ and $D_{0}\bar{D}^*$ molecular currents.
As a consequence, the mass obtained for the  $D_{0}\bar{D}^*$ 
molecular  current: $m_{D_{0}\bar{D}^*}=(4.96\pm 0.11)$  GeV is no 
longer compatible with the experimental mass of the meson $Y(4260)$.

\end{abstract}

\pacs{ 11.55.Hx, 12.38.Lg , 12.39.-x}
\maketitle

%

An error was made when calculating the spectral density of the
mixed condensate for the $D_{s0}(2317)\bar{D}_s^*(2110)$ molecular current,
 with $J^{PC}=1^{--}$, in ref.~\cite{alb}. Although the difference is
only a signal inone part of the mixed condensate, the result for the mass 
of the state changes significantly. The correct expression is
\beqa
&&\rho_1^{mix}(s)=-{m_c\mixs\over2^{7} \pi^4s}\int\limits_{\almi}^{\alma}
{d\al\over\alpha}
\int\limits_{\bemi}^{1-\al}{d\be\over\be^2}(2\alpha+\beta)\left[(\al+\be)m_c^2
-\al\be s\right].
\enqa

Doing the same analyses as done in ref.~\cite{alb} we arrive at
\beq
\lb{massmols}
m_{D_{s0}\bar{D}_s^*}=  (5.12\pm0.10)~\GeV~,
\enq
which is not in agreement with the $Y(4350)$ mass \cite{babar2,belle4} 
neither with the $Y(4660)$ mass \cite{belle4}. The above result was 
obtained using $5.5\leq\sqrt{s_0}\leq5.7~\GeV$. For values of 
$\sqrt{s_0}\leq5.4~\GeV$ we do not find an available Borel window.

For the molecular current $D_{0}\bar{D}^*$, using $5.4\leq\sqrt{s_0}\leq5.6~
\GeV$ we get
\beq
\lb{massmol}
m_{D_{0}\bar{D}^*}=  (4.96\pm0.11)~\GeV~,
\enq
which is much bigger than the mass of the meson $Y(4260)$ \cite{babar1}.

Therefore, we conclude that it is not possible to describe the $Y(4260)$ meson 
with a $D_{0}\bar{D}^*$ molecular current.


\end{document}